\documentclass[prl,aps,twocolumn,showpacs]{revtex4}
\include{graphics}
\usepackage{epsfig}
\usepackage{graphicx}
\usepackage{textcomp}

\begin{document}
\title{Spectral theory of metastability and extinction in birth-death systems}
\author{Michael Assaf and Baruch Meerson}
\affiliation{Racah Institute of Physics, Hebrew University of Jerusalem,
Jerusalem 91904, Israel} 

\begin{abstract}
We suggest a general spectral method for calculating statistics of multi-step
birth-death processes and chemical reactions of the type
$mA\to\hspace{-2.8mm}\hspace{2mm}nA$ ($m$ and $n$ are positive integers) which
possess an absorbing state. The method employs the generating-function formalism
in conjunction with the Sturm-Liouville theory of linear differential operators.
It yields accurate results for the extinction statistics and for the
quasi-stationary probability distribution, including large deviations, of the
metastable state. The power of the method is demonstrated on the example of
binary annihilation and triple branching $2A \to\hspace{-2.8mm}\hspace{2mm}
\emptyset,\,\,A \to\hspace{-2.8mm}\hspace{2mm}3A$, representative of the rather
general class of dissociation-recombination reactions.
\end{abstract}

\maketitle

Since the pioneering works of Delbr\"{u}ck \cite{delb},
Bartholomay \cite{barth} and McQuarrie \cite{McQuarrie}, kinetics of 
systems of birth-death type, containing a large but finite number of agents
(such as molecules, bacteria, cells, animals or even humans), have attracted
much attention in different areas of science and become a paradigm of theory of
stochastic processes \cite{gardiner,vankampen}. Birth-death models are
extensively discussed in chemistry, astrochemistry, epidemiology, population
biology, cell biochemistry, \textit{etc.} They are also well known in
non-equilibrium physics, and can be viewed in the context of reaction-limited
kinetics on a lattice, as opposed to more extensively studied diffusion-limited
kinetics \cite{Cardy}. While the behavior of the average number of particles in
such systems may be describable, at not too long times, by (mean-field)
\textit{rate equations}, fluctuations may lead to important quantitative or even
qualitative differences. This necessitates using the more general \textit{master
equation} which deals with the probability of having a certain number of
particles of each type at time $t$. The master equation is rarely solvable
analytically, and various approximations, often uncontrolled, are in use
\cite{gardiner,vankampen}, such as the Fokker-Planck equation which may suffice
unless one has to deal with large deviations or extinction phenomena
\cite{gaveau,kamenev,Sander}. Not much is known beyond the Fokker-Planck
description, though in particular cases the statistics were determined by using
approximations in the master equation
\cite{Sander,brau,nadler,dykman,laurenzi,Nasell,Biham} or, alternatively, by
introducing a generating function \cite{McQuarrie,gardiner,vankampen}, see
below, and developing different approximations in the equation describing its
evolution \cite{kamenev,escudero,assaf}.

In this work we advance the generating function technique by marrying it with
the Sturm-Liouville theory of linear differential operators.  This yields a
general and robust \textit{spectral} formalism, capable of providing accurate,
and often analytical, results for extreme statistics in a variety of (not
necessarily single-step) birth-death systems and chemical reactions. We
demonstrate the power of our method by a simple reaction of binary annihilation
and triple branching. An example of such a reaction is recombination of two
atoms $A$ and dissociation of the molecule $A_2$:  $A+A\to A_{2}$, and
$A_{2}+A\to 3A$, assuming that the $A_2$ molecules are always at hand
\cite{gates}. For H or N atoms this reaction occurs at high temperatures
\cite{byron}. We calculate the extinction probability as a function of time, the
mean time to extinction and the complete quasi-stationary probability
distribution of the long-lived metastable state, intrinsic to this problem.

\textit{Rate equation, master equation, generating function and spectral
theory.} Consider the binary annihilation and triple branching reactions $2A
\to\hspace{-4.3mm}^{\mu}\hspace{2mm} \emptyset$, and $A
\to\hspace{-4.3mm}^{\lambda}\hspace{2mm} 3A$ where $\mu,\lambda>0$ are rate
constants. The rate equation (or the mean field theory) of this simple system,
$dn/dt=2\lambda n-\mu n^2$, describes a nontrivial attracting steady state
$n_s=2\Omega$, where $\Omega=\lambda/\mu$. Fluctuations, caused by discreteness
of particles, invalidate this mean-field result owing to the existence of the
absorbing state $n=0$:  a state from which there is a zero probability of
exiting. At $\Omega\gg 1$, however, a long-lived (and therefore
quasi-stationary) fluctuating \textit{metastable} state is observed, once the
initial population is not too sparse. The statistics of the quasi-stationary
state and of the extinction times are the subjects of our interest here.

To account for discreteness of particles, we assume that the evolution of the
probability $P_n(t)$ to find $n$ particles at time $t$ is described, for $n>1$,
by the master equation
\begin{eqnarray}
\frac{d}{dt}{P}_{n}(t)&=&\frac{\mu}{2}\left[(n+2)(n+1)
P_{n+2}(t)-n(n-1)P_{n}(t)\right]\nonumber\\
&+&\lambda\left[(n-2)P_{n-2}(t)-nP_{n}(t)\right]\,. \label{mastereq}
\end{eqnarray}
Let us introduce the generating function \cite{McQuarrie,gardiner,vankampen}
\begin{equation}\label{generatingfun}
G(x,t)=\sum_{n=0}^{\infty}x^{n}P_{n}(t)\,,
\end{equation}
where $x$ is an auxiliary variable. $G(x,t)$ encodes all the probabilities:
\begin{equation}\label{probformula}
\left.P_{n}(t)=\frac{1}{n!}\frac{\partial^{n}G(x,t)}{\partial
x^{n}}\right|_{x=0}\,.
\end{equation}
Obviously, $G(x=1,t)=1$. Equations (\ref{mastereq}) and (\ref{generatingfun})
yield a partial differential equation (PDE) for $G(x,t)$:
\begin{equation}\label{partdiffeq}
\frac{\partial G}{\partial t} =
\frac{\mu}{2}(1-x^{2})\frac{\partial^{2} G}{\partial x^{2}}+\lambda
x(x^2-1)\frac{\partial G}{\partial x}\,.
\end{equation}
As the reaction we are dealing with conserves parity, $G(x,t)$ can be written as
\begin{equation}\label{generating}
G(x,t)=c_{1}G_{even}(x,t)+c_{2}G_{odd}(x,t)\,,
\end{equation}
where $c_{1}=\sum_{0}^{\infty}P_{2m}(t=0)$, and $c_{2}=1-c_{1}$. Therefore,
$G_{even}(x=\pm 1,t)=1$ and $G_{odd}(x=\pm 1,t)=\pm 1$. The steady
state solution of Eq. (\ref{partdiffeq}) 
is
\begin{equation}\label{stationary}
G_{st}(x,t)=c_{1}+c_{2}\frac{{\mbox{\it{erfi}}}\,(\sqrt{\Omega}x)}{\mbox{\it{erfi}}\,(\sqrt{\Omega})}\,,
\end{equation}
where $\mbox{\it{erfi}}(x)=(2/\sqrt{\pi})\int_{0}^{x}e^{t^2}dt$. Let the number
of particles at $t=0$ be \textit{even}: $n_0=2 k_0$, where $k_0$ is integer.  In
this case the parity conservation yields $c_1=1$ and $c_2=0$, so $G_{st}(x)=1$,
and the only true steady state is the empty state: $P_0=1$, while the rest of
$P_n$ are zero \cite{odd}.

To see how the population of $n_0=2 k_0$ particles at $t=0$ becomes extinct, we
introduce a new function $g(x,t)=G(x,t)-G_{st}(x)=G(x,t)-1$ which obeys
Eq.~(\ref{partdiffeq}) with homogenous boundary conditions $g(x=\pm 1,t)=0$.
Substituting $g(x,t)=e^{-\gamma t}\varphi(x)$, we obtain
\begin{equation}\label{orddiffeq}
(1-x^{2})\varphi^{\prime\prime}(x)+2 \Omega x(x^2-1)\varphi^{\prime}(x)+2
E\varphi(x)=0\,,
\end{equation}
where $E=\gamma/\mu$. 
Rewriting this ordinary differential equation in a self-adjoint form,
\begin{equation}\label{schroed}
\left[\varphi^{\prime}(x)\exp(-\Omega x^2)\right]^{\prime}+E w(x)
\varphi(x)=0\,,
\end{equation}
with the weight function $w(x)=2 \exp(-\Omega x^2)(1-x^2)^{-1}$,
we arrive at a standard eigenvalue problem of the Sturm-Liouville theory
\cite{Arfken}. Once we have found the complete set of orthogonal eigenfunctions
$\varphi_{k}(x)$ (which are all even), and the real eigenvalues $E_{k}$,
$k=1,2,\dots$, we can solve the time-dependent problem:
\begin{equation}\label{genseries}
G(x,t)= 1+\sum_{k=1}^{\infty} a_k \varphi_{k}(x) e^{-\mu E_{k}t}\,\,,
\end{equation}
where
\begin{equation}\label{an}
a_{k}=\frac{\int_{0}^{1}[G(x,t=0)-1]\varphi_{k}(x)w(x)dx}{\int_{0}^{1}\varphi_{k}^2(x)w(x)dx}\,,
\end{equation}
and $G(x,t=0)=x^{2k_0}$.

As all $E_k>0$, Eq. (\ref{genseries}) describes \textit{decay} of initially
populated states $k=1,2, \dots$, and the system approaches the empty state
$G(x,t\to \infty)=1$. We are interested in the case of $\Omega\gg 1$, where the
metastable state is expected to be long-lived. Elgart and Kamenev
\cite{kamenev1} showed that the eigenvalues $E_{2},E_{3},\dots$ scale like
${\cal O}(\Omega)\gg 1$. In contrast to these, the ``ground state" eigenvalue
$E_{1}$ is exponentially small, as will be proved \textit{a posteriori}. We will
be interested in sufficiently long times $\mu \Omega t=\lambda t\gg 1$, when the
contribution from the ``excited" states to $G(x,t)$ becomes negligible, and we
can write
\begin{equation}\label{genseriesapprox}
G(x,t)= 1+a_1 \,\varphi_{1}(x)\,e^{-\mu E_{1}t}\,.
\end{equation}
Let us proceed to the ground state calculations.

\textit{Ground state calculations.} As $\varphi_1(x)\equiv\varphi(x)$ is an even
function, it suffices to consider the interval $0\le x\le 1$. We will employ the
strong inequality $\Omega \gg 1$ and find the (very small) eigenvalue $E_1$ and
the corresponding eigenfunction of Eq.~(\ref{orddiffeq}) by a matched asymptotic
expansion, see \textit{e.g.} Ref. \cite{orszag}. In most of the region $0\le
x<1$ (the bulk) we can treat the last term in Eq. (\ref{orddiffeq})
perturbatively. In the zero order we put $E_1=0$ and arrive at the
\textit{steady state} equation $\varphi_{b}^{\prime\prime}(x)-2 \Omega
x\varphi_{b}^{\prime}(x)=0$,
whose even solution is $\varphi_{b}^{(0)}=1$. 
Now we put $\varphi_{b}(x)=1+\delta\varphi_{b}(x)$, where
$\delta\varphi_{b}(x) \ll 1$, and obtain in the first order
\begin{equation}\label{orddiff1}
\delta\varphi_{b}^{\prime\prime}(x)-2 \Omega x\delta\varphi_{b}^{\prime}(x)=-2
E_{1}(1-x^2)^{-1}\,.
\end{equation}
The solution for $\varphi_b(x)$ takes the form:
\begin{equation}\label{phisol}
\varphi_{b}(x)= 1-2E_{1}\int_{0}^{x}e^{\Omega
s^2}ds\int_{0}^{s}\frac{e^{-\Omega r^2}}{1-r^{2}}dr\,.
\end{equation}
As $\Omega\gg 1$, we can omit the $r^2$ term in the denominator of the inner
integral in Eq. (\ref{phisol}) (this omission is illegitimate too close to
$x=1$, but the bulk solution is invalid there anyway, see below). We obtain
\begin{eqnarray}
\varphi_{b}(x)&\simeq& 1-2E_{1}\int_{0}^{x}e^{\Omega
s^2}ds\int_{0}^{s}e^{-\Omega r^2}dr\nonumber\\
&=& 1-E_{1}x^2\;_{2}F_{2}\left(1,1\,;\frac{3}{2},2\,;\Omega x^2\right)\,,
\label{phisolapprox}
\end{eqnarray}
where $_{2}F_{2}(a_1,a_2;b_1,b_2;x)$ is the generalized hypergeometric function
\cite{Abramowitz}, while $E_1$ is yet unknown. It is easy to check that the bulk
solution is valid [$\delta\varphi_{b}(x)\ll 1$] as long as $1-x\gg 1/\Omega$.

In the boundary layer $1-x\ll 1$ we can again disregard, at $\Omega\gg 1$,  the
(exponentially small) last term in Eq.~(\ref{orddiffeq}).
The resulting equation is again $\varphi_{l}^{\prime\prime}(x)-2\Omega
x\varphi_{l}^{\prime}(x) =0$.
Its non-trivial solution, obeying the boundary condition at $x=1$, is
\begin{equation}\label{phiwallapprox}
\varphi_{l}(x)= C\int_x^1 e^{\Omega s^2}ds\simeq
C\frac{e^{\Omega}}{2\Omega}\left[1-e^{-\Omega(1-x^2)}\right]\,,
\end{equation}
where $C$ is a yet unknown constant.

Now we demand that, in the common region $1/\Omega\ll 1-x\ll 1$, the
proper asymptote of the bulk solution (\ref{phisolapprox}), obtained
by moving to infinity the upper limit in the inner integral of Eq.
(\ref{phisolapprox}):
\begin{equation}\label{phisolapprox1}
\varphi_{b}(x)\simeq
1-\frac{\sqrt{\pi}E_{1}}{\sqrt{\Omega}}\int_{0}^{x}e^{\Omega
s^2}ds\simeq 1-\frac{\sqrt{\pi}E_{1}}{2\Omega^{3/2}}e^{\Omega
x^{2}}\,,
\end{equation}
coincides with the boundary layer solution (\ref{phiwallapprox}). Equations
(\ref{phiwallapprox}) and (\ref{phisolapprox1}) yield
\begin{equation}\label{results}
E_{1}=\frac{2 \Omega^{3/2}}{\sqrt{\pi}}\,
e^{-\Omega}\;\;\;\mbox{and}\;\;\;C=2\Omega e^{-\Omega}\,.
\end{equation}
As expected, the lowest eigenvalue $E_{1}$ is exponentially small in $\Omega$.
The respective eigenfunction is
\begin{equation}
\varphi (x)\simeq\left\{\begin{array}{ll}
\varphi_{b}(x)=1-E_{1}x^2\;_{2}F_{2}\left(1,1;\,\frac{3}{2},2\,;\Omega
x^2\right)\nonumber\\
\;\;\;\;\;\mbox{for}\; 1-x^2\gg 1/\Omega\,,\nonumber\\
\\
\varphi_{l}(x)=1- e^{-\Omega (1-x^2)}\nonumber\\
\;\;\;\;\;\mbox{for} \; 1-x^2 \ll 1\,.\end{array}\right.
\label{eigenfunction}
\end{equation}
Now we use Eq.~(\ref{an}) to calculate the coefficient $a_{1}$ entering
Eq.~(\ref{genseriesapprox}). While evaluating the integrals, we notice that the
main contributions come from the bulk region $1-x^2 \gg 1/\Omega$, and it
suffices to take the eigenfunction $\varphi_b(x)$ in the zeroth order, that is
$\varphi_b^{(0)}(x)\simeq 1$. Evaluating the integral in the nominator of
Eq.~(\ref{an}), we notice that, the term $x^{2k_0}$ under the integral is
negligible compared to $1$. As a result, the integrals in the nominator and
denominator become identical up to a minus sign. Therefore, $a_{1}\simeq -1$
which completes our solution (\ref{genseriesapprox}).

\begin{figure}[h]
\includegraphics[width=7cm,clip=]{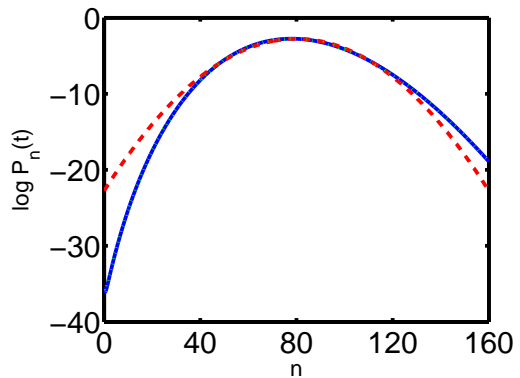}
\caption{(Color online) The natural logarithms of the probability distribution
(\ref{prob}) (the blue solid line), of its large-$n$ asymptotics
(\ref{probapprox}) (the green dotted line), and of the normal distribution with
$\bar{n}$ and $\sigma$ from Eqs.~(\ref{nbar}) and (\ref{nvar}) (the red dashed
line), for $\Omega=40$ and $\mu E_1 t \ll 1$.} \label{probfig}
\end{figure}

\textit{Statistics of the quasi-stationary state.} We start this part with
calculating the average number of particles $\bar{n}$ and the standard deviation
$\sigma$ at intermediate times $\Omega^{-1}\ll\mu t\ll E_{1}^{-1}$. Using
Eq.~(\ref{genseriesapprox}), we obtain
\begin{equation}\label{nbar}
    \bar{n}=\left.\partial_x G\right|_{x=1}= 2\Omega\,,
\end{equation}
which coincides with the mean field result. Furthermore,
\begin{equation}\label{nvar}
   \sigma^2=\bar{n^2}-\bar{n}^2=
\left. \left[\partial_{xx}^2 G+\partial_x G -(\partial_x
G)^2\right]\right|_{x=1} = 4\Omega\,.
\end{equation}
where we have used for $\varphi (x)$ its boundary layer asymptote
$\varphi_{l}(x)$ from  Eq.~(\ref{eigenfunction}). One can see that, at
intermediate times $\Omega^{-1}\ll\mu t\ll E_{1}^{-1}$, the system stays in the
(weakly fluctuating) quasi-stationary state. What is the complete probability
distribution $P_{n}(t)$ of the quasi-stationary state at these times? For $n=0$
we obtain
\begin{equation}\label{p0}
P_0(t)=G(x=0,t)= 1 - e^{-\mu E_{1} t}
\end{equation}
which, at $\mu E_{1} t\ll 1$, is very small. For (even) nonzero values of $n$,
Eqs.~(\ref{probformula}) and (\ref{genseriesapprox}) yield
\begin{equation}\label{prob}
P_{n}(t)= \frac{2E_{1} (4\Omega)^{n/2-1}(n/2-1)!}{n!}\,e^{-\mu E_{1} t}\,.
\end{equation}
For $n\gg 1$ we can use Stirling's formula and obtain
\begin{equation}\label{probapprox}
P_{n}(t)\simeq \frac{E_{1}}{\sqrt{2}n\Omega} e^{\frac{n}{2}\left(1+\ln
\frac{2\Omega}{n}\right)-\mu E_{1} t}\,.
\end{equation}
Notably, \textit{all} of the probabilities $P_{n}(t)$ ($n>0$) decay with time,
while $P_{0}(t)$ grows. One can check  that the most probable state coincides
with $\bar{n}= 2\Omega$. In the vicinity of $n=\bar{n}$, $P_{n}(t)$ from
Eq.~(\ref{probapprox}) can be approximated by a normal distribution with the
mean $\bar{n}$ and standard deviation $\sigma$, given by Eqs.~(\ref{nbar}) and
(\ref{nvar}), respectively. The tails of the true distribution, however, are
strongly non-Gaussian. A comparison between our analytic result (\ref{prob}),
the large-$n$ approximation (\ref{probapprox}), and the normal distribution is
shown in Fig. \ref{probfig}. One can see that Eq.~(\ref{probapprox}) is very
accurate, whereas the gaussian approximation strongly overpopulates the low-$n$
tail and underpopulates the high-$n$ tail.

\begin{figure}[ht]
\includegraphics[width=7cm,clip=]{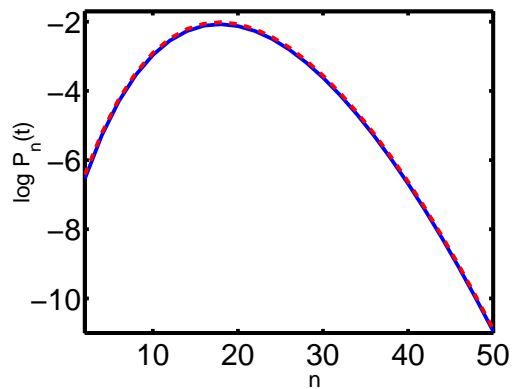}
\caption{(Color online) The red dashed line: the natural logarithm of the
probability distribution (\ref{prob}) at $\mu E_1 t \ll 1$. The blue solid line:
a numerical solution of the master equation~(\ref{mastereq}), see text.
The parameters are $\Omega=10$, 
$n_0=2k_0=40$.} \label{probmaster}
\end{figure}

Figure \ref{probmaster} compares our analytic result (\ref{prob}) with a
numerical solution of the (truncated) master equation (\ref{mastereq}) with
$(d/dt)P_n(t)$ replaced by zeros and $P_0=0$. The two curves are almost
indistinguishable for $\Omega=10$. In fact, good agreement is observed already
for $\Omega={\cal O}(1)$, and it rapidly improves further as $\Omega$ increases.

\begin{figure}[ht]
\includegraphics[width=7cm,clip=]{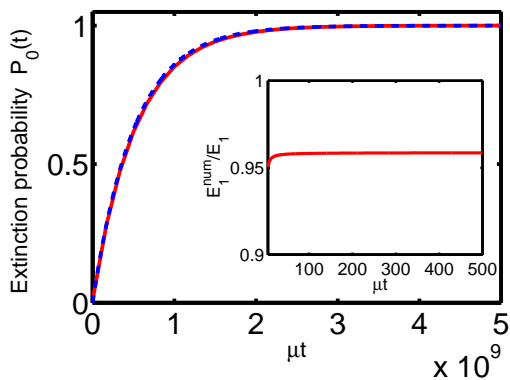}
\caption{(Color online) Extinction probability $P_{0}(t)$ from Eq.~(\ref{p0})
(the blue dashed line) and from a numerical solution of Eq. (\ref{partdiffeq})
(the red solid line) for $\Omega=25$ and $n_0=2k_0=100$. The inset shows,  at
times $\mu t\gg 1/\Omega$, the ratio of the numerical ground state eigenvalue
$E_1^{num}=-\log [1-G(0,t)]/\mu$ and the analytical one, given by Eq.
(\ref{results}). The deviation is less than 4\%, that is within error ${\cal
O}(1/\Omega)$.} \label{numexact}
\end{figure}

\textit{Extinction time statistics.} The quantity $P_0(t)$ is the probability of
extinction at time $t$. The extinction probability density is $p(t)=dP_0(t)/dt$.
Using Eq. (\ref{p0}), we obtain an exponential distribution:
\begin{equation}\label{ltdist}
p(t)\simeq \mu E_{1} e^{-\mu E_{1} t}\;\;\;\mbox{at}\;\;\;\lambda t\gg 1\,.
\end{equation}
The average time to extinction, $\bar{\tau} = \int_0^{\infty}t
p(t)\,dt\simeq(\mu E_{1})^{-1}$,
is exponentially large, at $\Omega\gg1$, as expected. Figure \ref{numexact}
compares the analytical result (\ref{p0}) for $P_{0}(t)$ with $G(0,t)$ found by
solving Eq.~(\ref{partdiffeq}) numerically with the boundary conditions $G(\pm
1,t)=1$ and the initial condition $G(x,t=0)=x^{2k_0}$. The inset compares the
analytical and numerical ground state eigenvalues, and good agreement is
observed.

\textit{Final comments.} The spectral formalism yields accurate extinction time
statistics and a complete quasi-stationary probability distribution of the
metastable state for a wide class of birth-death processes which possess an
absorbing state and are describable by a master equation. In this formalism, the
problem of computing these statistics is reduced to a problem (familiar to every
physicist) of finding a ground-state eigenfunction and eigenvalue of a linear
differential operator. We have demonstrated the formalism by an example of
binary annihilation and triple branching, but the formalism is general and can
be used for a variety of kinetics. In most interesting cases of
\textit{long-lived} metastable states, a large parameter (the average number of
particles in the metastable state) is always present in the problem. This paves
the way to a perturbative treatment, like in the example we have considered.

The spectral formalism should be also efficient when the absorbing state is at
infinity, rather than at zero. For systems of this type the rate equation yields
a stable non-empty steady state, but an account of fluctuations brings about an
unlimited population growth, see \textit{e.g.} Ref. \cite{kamenev}. In that case
the ground-state eigenvalue is expected to be \textit{negative}. Finally, the
use of spectral formalism is not at all limited to systems possessing an
absorbing state \cite{assaf}.

We acknowledge useful discussions with Alex Kamenev and Vlad Elgart and thank
Len Sander and Uri Asaf for advice. The work was supported by the Israel Science
Foundation (grant No. 107/05) and by the German-Israel Foundation for Scientific
Research and Development (Grant I-795-166.10/2003).


\begin{thebibliography}{99}
\bibitem{delb} M. Delbr\"{u}ck, J. Chem. Phys. \textbf{8}, 120 (1940).
\bibitem{barth} A.F. Bartholomay, Bull. Math. Biophys. \textbf{20}, 175
(1958).
\bibitem{McQuarrie} D.A. McQuarrie,  J. Appl. Prob. \textbf{4}, 413 (1967).
\bibitem{gardiner} C.W. Gardiner, \textit{Handbook of Stochastic
Methods} (Springer, Berlin, 2004).
\bibitem{vankampen} N.G. van Kampen, \textit{Stochastic Processes in Physics
and Chemistry} (North-Holland, Amsterdam, 2001).
\bibitem{Cardy} J.L. Cardy and U.C. T\"{a}uber, J. Stat. Phys. \textbf{90} (1-2), 1
(1998); U.C. T\"{a}uber, M. Howard and B.P. Vollmayr-Lee, J. Phys. A
\textbf{38}, R79 (2005).
\bibitem{gaveau} B. Gaveau, M. Moreau, and J. Toth, Lett. Math. Phys. \textbf{37}, 285
(1996).
\bibitem{kamenev} V. Elgart and A. Kamenev, Phys. Rev. E
\textbf{70}, 41106 (2004).
\bibitem{Sander} C.R. Doering, K.V. Sargsyan, and L.M. Sander, Multiscale Model. and
Simul. \textbf{3}, 283 (2005), and references therein.
\bibitem{brau} C.A. Brau, J. Chem. Phys. \textbf{47}, 1153
(1967).
\bibitem{nadler} W. Nadler and K. Schulten, J. Chem. Phys. \textbf{82}, 151 (1985);
Z. Phys. B \textbf{59}, 53 (1985).
\bibitem{dykman} M.I. Dykman, E. Mori, J. Ross, and P.M. Hunt, J. Chem. Phys.
\textbf{100}, 5735 (1994).
\bibitem{laurenzi} I.J. Laurenzi, J. Chem. Phys. \textbf{113}, 3315
(2000).
\bibitem{Nasell} I. Nasell, J. Theor. Biol. \textbf{211}, 11
(2001).
\bibitem{Biham} O. Biham, I. Furman, V. Pirronello, and G. Vidali, Astrophys.
J. \textbf{553}, 595 (2001).
\bibitem{escudero} C. Escudero, Phys. Rev. E \textbf{74}, 010103(R)
(2006).
\bibitem{assaf} M. Assaf and B. Meerson, Phys. Rev. E \textbf{74}, 041115 (2006).
\bibitem{gates} B.C. Gates, J.R. Katzer and G.C.A. Schuit,
\textit{Chemistry of Catalytic Processes} (McGraw-Hill, New York, 1979).
\bibitem{byron} S.R. Byron, J. Chem. Phys. \textbf{30}, 1380
(1959); \textbf{44}, 1378 (1966).
\bibitem{odd} For an \textit{odd} number of particles there is no extinction. Here
$c_1=0$ and $c_2=1$, and one obtains a true steady state:
$P_{n}=(4\Omega)^{n/2}\Gamma(n/2)/[\pi n!\,\mbox{\it{erfi}}(\sqrt{\Omega})]$,
$n=1,3,5,\dots.$
\bibitem{Arfken} G. B. Arfken, \textit{Mathematical Methods for
Physicists} (Academic Press, London, 1985).
\bibitem{kamenev1} V. Elgart and A. Kamenev (private communication).
\bibitem{orszag} C.M. Bender and S.A. Orszag, \textit{Advanced
Mathematical Methods for Scientists and Engineers} (Springer, New York, 1999).
\bibitem{Abramowitz} M. Abramowitz, \textit{Handbook of
Mathematical Functions} (National Bureau of Standards, Washington, 1964).


\end{thebibliography}
\end{document}